\documentclass[structabstract]{aa}  
\usepackage{graphicx}
\usepackage[]{natbib}
\usepackage{txfonts}
\begin{document}

        \title{Exploring jet-launching conditions for SFXTs}
    \titlerunning{Exploring jet-launching conditions for SFXTs}
    \authorrunning{F. Garc\'ia, D. N. Aguilera, G. E. Romero}
%   \subtitle{...}
    \author{Federico Garc\'ia
          \inst{1,2}\thanks{Fellow of CONICET, Argentina.}, Deborah N. Aguilera \inst{3,4}
          \and
          Gustavo E. Romero\inst{1,2}
          }

   \institute{Instituto Argentino de Radioastronom\'{\i}a (CCT La Plata, CONICET), C.C.5, (1894) Villa Elisa, Buenos Aires, Argentina.\\
                 \email{[fgarcia,romero]@iar-conicet.gov.ar}
                 \and
Facultad de Ciencias Astron\'omicas y Geof\'{\i}sicas, Universidad Nacional de La Plata, Paseo del Bosque, B1900FWA La Plata, Argentina.
         \and
             Laboratorio Tandar, CNEA, Av. Gral. Paz 1499, 1430 San Mart\'in, Buenos Aires, Argentina \& CONICET.\\
             \email{aguilera@cnea.tandar.gov.ar}
             \and
             German Aerospace Center, Institute for Space Systems, Robert Hooke Str. 7, 28359 Bremen, Germany                           
             }

        \offprints{F. Garc\'{\i}a}  

   \date{Received ; accepted }

% \abstract{}{}{}{}{} 
% 5 {} token are mandatory
 
  \abstract
   {In the magneto-centrifugal mechanism for jet formation, accreting neutron stars are assumed to produce relativistic jets only if their surface magnetic field is weak enough ($B\sim10^8$~G). However, the most common manifestation of neutron stars are pulsars, whose magnetic field distribution peaks at $B\sim10^{12}$~G. If the neutron star magnetic field has at least this strength at birth, it must decay considerably before jets can be launched in binary systems.}
   {We study the magnetic field evolution of a neutron star that
accretes matter from the wind of a high-mass stellar companion
so that we can  constrain the accretion rate and the impurities in the crust, which are necessary conditions for jet formation. }
   {We solved the induction equation for the diffusion and convection of the neutron star magnetic field confined to the crust, assuming spherical accretion in a simpliflied one-dimensional treatment. We incorporated state-of-the-art microphysics, including consistent thermal evolution profiles, and assumed two different neutron star cooling scenarios based on the superfluidity conditions at the core.}
   {We find that in this scenario, magnetic field decay at long timescales is governed mainly by the accretion rate, while the impurity content and thermal evolution of the neutron star play a secondary role. For accretion rates \mbox{$\dot{M}\gtrsim 10^{-10}$~M$_{\odot}$~yr$^{-1}$}, surface magnetic fields can decay up to four orders of magnitude in $\sim$10$^7$~yr, which is the timescale imposed by the evolution of the high-mass stellar companion in these systems. Based on these results, we discuss the possibility of transient jet-launching in strong wind-accreting high-mass binary systems like supergiant fast X-ray transients.}  
   {}
   \keywords{Stars: neutron, magnetic fields, accretion, X-rays: binaries}
   \maketitle
%
%________________________________________________________________

\section{Introduction}

A new class of short X-ray transients was discovered by {\it INTEGRAL} observations of the Galactic plane. The so-called supergiant fast X-ray transients (SFXTs) are presumably composed of a compact object (a neutron star --NS-- or a black hole) and an OB supergiant. These binaries show short flares that last from a few hours to days, reaching $L_{\rm X}\sim 10^{36}-10^{37}$~erg~s$^{-1}$ and returning to quiescent levels of $L_{\rm X}\sim 10^{32}$~erg~s$^{-1}$ \citep{sguera2005,negueruela2006}. They are more frequently in an intermediate state in which $L_{\rm X}\sim10^{33}-10^{34}$~erg~s$^{-1}$, because of residual accretion onto the compact object \citep{sidoli2008}.

At present, the SFXTs class has ten members, identified through the association of the transient X-ray source with blue supergiant companions. In addition, there are  several candidates without confirmed optical/IR counterparts \citep[see e.g.,][for a recent review]{sidoli2011}. In at least four SFXTs, the discovery of X-ray pulsations that reach from 4.7 to 228~s confirms the presence of a NS. Orbital periods have also been measured; they range from 3.3 to 165 days. Three models have been proposed for the fast-flaring mechanism \citep[see the review by][]{sidoli2009}: (i) spherically symmetric clumpy winds, where the X-ray flares are produced when a dense clump is accreted by the compact object \citep{intzand2005,walterzurita2007,negueruela2008,ducci2009}; (ii) anisotropic winds, where the flare occurs when the compact object crosses a slow and dense equatorial wind component, which
increases the accretion process \citep{sidoli2007a}; (iii) gated mechanisms, where accretion is inhibited by a centrifugal or magnetic barrier \citep{grebenevsunyaev2007,bozzo2008}, which
requires an NS with a strong magnetic field ($B \sim 10^{14}-10^{15}$~G) and a slow spin period ($P_{\rm s} \sim 10^3$~s). However, magnetar activity in SFXTs has not been detected yet. An estimate of a low magnetic field ($B \sim 10^{11}$~G) has been obtained through a cyclotron line in the spectrum of SFXT IGR J18483--0311, for an electron origin. However, a magnetar cannot be discarded if this cyclotron line is caused by protons, for which $B \sim 5\times 10^{14}$~G.

Four unidentified $\gamma$-ray transient sources, AGL J2022+3622, 3EG J1837--0423, 3EG J1122--5946, and AGL J1734--3310, are spatially correlated with three SFXTs (or candidates): IGR J20188+3647, AX J1841.0--0536, IGR J11215--5952 \citep{sguera2009}, and one intermediate SFXT: IGR J17354--3255 \citep{sguera2011}. These possible associations raise the possibility that SFXT might also produce radiation at energies above 100 MeV. Recently, \citet{sgueraromero2009} developed a model for the $\gamma$-ray emission from SFXT AX J1841.0--0536 based on the hypothesis that the high-energy radiation is originated from the cooling of relativistic particles accelerated in a collimated outflow from the NS. Although jets have been confirmed in some NSs that belong to low-mass X-ray binaries (LMXBs), in high-mass X-ray binaries (HMXBs) that host an NS no jet has been detected so far, although currently a possible
jet origin is discussed for  HMXB \mbox{LS I +61 303}   for its $\gamma$-ray emission \citep{romero2007}.

The magneto-centrifugal mechanism for jet formation in accreting NSs requires the accreted material to drag magnetic field lines near to the surface of the compact object. The fluid pressure in this region must be higher than that exerted by the NS magnetic field. If this is the case, the field should not be greater than $B\sim10^8$~G \citep{massikauf2008}. For LMXBs with old NSs ($\gtrsim$10$^9$~yr), the magnetic field has enough time to decay several orders of magnitude from a typical value $\sim$10$^{12}$~G of the initial field to the required field for jet launching. On the other hand, the timescales are quite short for HMXBs, since the donor star has a short lifetime as well ($\sim$10$^7$~yr). However, this type of stars presents strong and highly inhomogeneous winds \citep{runacres2005,owocki2006, negueruela2010} that enhance
the accretion process onto the NS and might trigger an accelerated magnetic field decay at the NS surface by advecting the currents originating the magnetic field to the interior of the star in a burial process \citep[see the pioneering work from][]{bisno1974}. Moreover, impurities in the crust due to accretion might amplify its resistivity.

In this work we explore necessary conditions for an NS that belongs to an HMXB system to undergo a surface magnetic field decay from an initial field of the order of $B \sim 10^{12}$~G to a final value of $B\lesssim10^8$~G, which could allow for jet formation in the system. We used a one-dimensional spherical accreting NS model, including state-of-the-art mycrophysics, electrical conductivity, and thermal evolution profiles, considering qualitatively different scenarios in terms of accretion rates, superfluidity in the NS core, and impurity content in the NS crust, for which we numerically solved the induction equation for the magnetic field evolution with appropriate initial and boundary conditions on a timescale of $10^7$~yr. With this first approach to the problem, we aim to put initial constraints in preparation for deeper investigations that use a more realistic model for the magnetic field evolution \citep[see][]{pons2012}.

The structure of this paper is as follows: in Sect.~2 we present the accreted NS crust model, the thermal profiles, and the model adopted for the magnetic field evolution according to the induction equation and transport properties in the NS crust. In Sect.~3 we show our numerical results, and in Sect.~4 we discuss them in the context of the SFXTs. Finally, we summarize our conclusions in Sect.~5.

\section{Underlying neutron star model}

\subsection{Accreted neutron star crust}
\label{accretedNScrust}

To study the magnetic field evolution, we first constructed a background NS model using an equation of state to represent both the NS crust and its core, based on the effective nuclear interaction SLy \citep{douchin2001}, and considering a modified crust composition due to the accreted material \citep{haensel2008}. At low densities in the outer crust ($\rho \sim 10^{10}$~gr~cm$^{-3}$) the altered crust composition presents nuclei with atomic number $Z<20$, while in catalyzed matter $Z$ is 40--50. At intermediate densities, the lattice
presents nuclei with mass number $A\la100$ in contrast to the
$A\sim300$ that is typical of isolated NSs. At high densities in the inner crust ($\rho > 10^{13}$~gr~cm$^{-3}$), the accreted matter composition is very similar to the non-accreted case because
of free neutron gas that scatters on the ion lattice, and thus the effect caused by accretion become less important than in the outer regions.

To investigate whether possible differences might arise from the NS internal structure, we propose two configurations: a $1.4$~M$_{\odot}$ NS, which we call low-mass (LM) model, and an NS of $1.8$~M$_{\odot}$, named high-mass (HM) model. In Table \ref{NSmasses} we list the most important properties of these two models.   

%%%%%%%%%%%%%%%%%%%%%%%%%%%%%%%%%%%%%%

\begin{table}
\caption{Central density $\rho_{\rm c}$, stellar mass $M$, stellar radius $R_{\rm NS}$ , and 
crust thickness $\Delta R_{\rm crust}$ for the low-mass (LM) and high-mass (HM) NS models. }
\label{NSmasses}
\centering
\begin{tabular}{c c c c c}
\hline \hline
 Neutron Star &$\rho_{\rm c}$ &  $M$  & $R_{\rm NS}$ & $\Delta R_{\rm crust}$ \\
  Model     &~~(g~cm$^{-3}$)~~ &~~($M_{\odot}$)~~& ~~(km)~~ & ~~(km)~~  \\
\hline
Low Mass (LM) &$9.9\times10^{14} $&  $1.4$& $11.72$& $0.93$\\
High Mass (HM) &$1.4\times10^{15} $&  $1.8$& $11.34$& $0.59$\\
\hline
\end{tabular}
\end{table}
%%%%%%%%%%%%%%%%%%%%%%%%%%%%%%%%%%%%%%%%

For the chosen equation of state, the crust-core interface is at 
$\rho_{\rm CC} = 0.46\, \rho_0$, where $\rho_0=2.8 \times 10^{14}$ g~cm$^{-3}$ is the
nuclear saturation density. The crust thickness, defined as the distance from this crust-core interface to the NS surface, is about $\sim$1~km in the LM model, which is $\sim$60\% thinner in the HM model. This thickness represents a characteristic length scale for the confinement of the crustal magnetic field, which is supported by the currents in the crust.

\begin{figure}
  \centering
  \includegraphics[width=5cm,angle=-90]{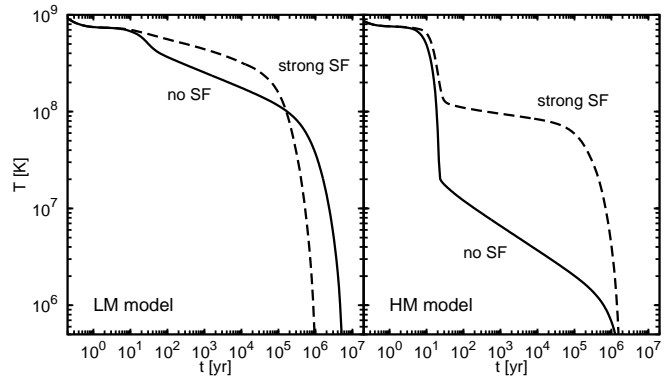}
  \caption{Thermal profiles adopted from \citet{aguilera2008}. Left (right) panel shows the thermal evolution for
    two extreme cases for the superfluidity in the core of the LM (HM) model.}
  \label{cooling}
\end{figure}

\subsection{Thermal profiles}
\label{thermal_profiles}

We assumed an isothermal crust with a temperature that evolves following thermal profiles shown in Figure \ref{cooling}, taken from \citet{aguilera2008}. 

To analyse the qualitative effects of thermal profiles on the magnetic field evolution, we incorporated two extreme cases for neutron superfluidity (SF) in the p-wave ($^3P_2$ state) in the NS core: strong superfluidity ({\it strong SF} model), where we considered very high critical temperatures, $T_c \approx 6 \times 10^9$~K, and no superfluidity ({\it no SF} model) for this nucleon. We considered in both models an s-wave superfluidity of neutrons in the crust and protons in the core (with $T_c \approx 8 \times 10^9$~K, $\approx 6 \times 10^9$~K, respectively), but these correlations leave a much weaker imprint on the cooling curves than the neutrons in the core. 

For $t \lesssim 10^5$~yr, the neutron pairing in the p-wave deaccelerates the cooling compared with the non-superfluid case. In this first stage, the more massive star suffers a faster cooling due to the efficient emission of neutrinos through the direct Urca process. When the temperature is low enough (at, $t>10^5$~yr), photon emission dominates the cooling and the suppresion of specific heat by superfluidity gives as result an accelerated cooling, with final temperatures $T < 10^5$~K on timescales $t < 10^7$~yr.

\subsection{Magnetic field evolution}

Under the assumptions of high electrical conductivity and non-relativistic fluid velocities, magnetic field evolution in the NS crust is governed by the magnetohydrodynamic induction equation: 
 
\begin{equation}
\frac{\partial {\mathbf B}}{\partial t} = - \frac{c^2}{4\pi}
{\mathbf \nabla} \times \left(\frac{1}{\sigma} {\mathbf \nabla} \times {\mathbf
  B}\right) +
{\mathbf \nabla} \times \left({\mathbf v} \times {\mathbf B}\right) , 
\label{eq:ind}
\end{equation}

\noindent where $c$ is the light velocity, ${\mathbf v}$ the velocity of the fluid, and $\sigma$ the electrical conductivity. 

The partial differential equation (Eq.~\ref{eq:ind}) for the evolution of the magnetic field ${\mathbf B}$ involves two terms. The first one is known as the diffusion term, which essentially
depends on the electrical conductivity. The second one is the advective term, which originates from the magnetic field line dragging caused by the movement of the fluid that holds the sources.

Assuming a dipolar magnetic field configuration, the vector potential reads ${\mathbf A}=(0,0,A_\phi)$, where $A_\phi=s(r,t)\sin \theta /r$. Here $r$ is the radial coordinate and $\theta$ and $\phi$
are the polar and azimuthal angles. Then, assuming a spherical inflow for the accreting matter, the induction equation reduces to a one-dimensional partial differential equation for the Stokes stream function, $s(r,t)$:

\begin{equation}
\frac{\partial {s}}{\partial t} =
\frac{c^{2}}{4\pi\sigma}\left(\frac{\partial ^{2} s}{\partial
  r^{2}}-\frac{2s}{r^2}\right) - v_r\frac{\partial s}{\partial r} .
\label{eq:stokes1}
\end{equation}

\noindent The fluid velocity, $v_r(r)$, is obtained by the continuity condition from the accretion rate,
$\dot{M}$, which is a free parameter in this model:

\begin{equation}
v_r(r) = -\frac{\dot{M}}{4 \pi r^2 \rho(r)}
\label{veloc_eq}
,\end{equation}

\noindent where $\rho(r)$ is the density profile of the accreted NS crust. It is important to note that for the accretion rates and magnetic field strengths considered here, columnar accretion at the NS magnetic poles is expected, involving a small area of the NS surface. This certainly motivates a two-dimensional treatment of the incoming matter flow and its effect on the accretion area, which is beyond the scope of the present work. We instead focused on qualitative aspects of the magnetic field decay that are expected for the columnar accretion model as well.

To solve the differential equation (\ref{eq:stokes1}), it is necessary to assume proper boundary and initial conditions for the problem. At the surface, the boundary condition originates from matching of the internal geometry with a vacuum dipolar exterior solution. Following \citet{kb1997}, the interior boundary condition is obtained assuming that the electrical conductivity at the NS core is several orders of magnitude higher than in the NS crust. For the initial condition, we assumed that the
currents that hold the magnetic field are expelled from the core during the NS formation by superconducting protons, and therefore the magnetic field is confined to the crust \citep{gu1994}:

\begin{equation}
s(r,t=0)= \left\{
 \begin{array}{c l l}
0 & \textrm{if} & r < r_{\rm i}\\
\left(\frac{r-r_{\rm i}}{R_{\rm NS}-r_{\rm i}}\right)^2 & \textrm{if} & r_{\rm i}< r \leq R_{\rm NS}
\end{array}
\right. ,
\label{cond_ini}
\end{equation}

\noindent where $r_{\rm i}$ is the radial distance for the initial expulsion of the magnetic field, and we normalized $s(R_{\rm NS},t=0)=1$, for simplicity. Under these assumptions, the magnetic field at the surface of the NS at any time can be obtained by means of the Stokes function, $B(R_{\rm NS},t) = s(R_{\rm NS},t) \cdot B(R_{\rm NS},t=0)$, where $B(R_{\rm NS},t=0)$ is the initial surface magnetic field that we fixed to the canonical value $10^{12}$~G.

Electrical and thermal conductivity at the NS crust play a fundamental role in both the magnetic and thermal
evolution. In particular, the electrical conductivity governs magnetic field evolution through the diffusion term of the induction equation (\ref{eq:ind}). In the NS crust, the principal charge carriers are the electrons, and thus the transport coefficient, $\sigma$, is determined taking into account every electron scattering-on processes: ions (p), phonons (ph), and impurities (Q). We used the non-quantizing electron conductivity from a public code\footnote{http://www.ioffe.ru/astro/conduct/index.html} based on \cite{potekhin1999}, with the modifications introduced by \cite{chugunov2012}. These routines give $\sigma$ as a function of the temperature, $T$, density, $\rho$, magnetic field, $B$, atomic and mass numbers, ($Z$, $A$), auxiliary mass number, which incorporates the free neutron gas, $A^*$, and the impurity content parameter, $Q=Z_{\rm imp}^2$.

The impurity content is defined as a measure of the charge dispersion in the lattice, $Q = \frac{1}{n_i}\sum_{n'} n' (Z-Z')^2$, where $n_i$ is the number density of the dominant ion of charge $Z$ and $n'$ are the number densities of the interloper species of charge $Z'$. The conductivity associated with $Q$ dominates at low temperatures, when the conductivity due to scattering-on phonons becomes irrelevant. We assumed two extreme values for the impurity parameter in the outer crust: $Q=0.1$, representing an almost perfect crystal, and $Q=5$, for a high impurity level. In the inner crust, beyond the neutron drip (ND) density ($\rho > \rho_{\rm ND} = 4.2 \times 10^{11}$~g~cm$^{-3}$), a free neutron gas that scatters on the nuclei of the lattice prevents the formation of impurities, and therefore we set $Q=0$.

\begin{figure}
  \centering
  \includegraphics[width=6cm,angle=-90]{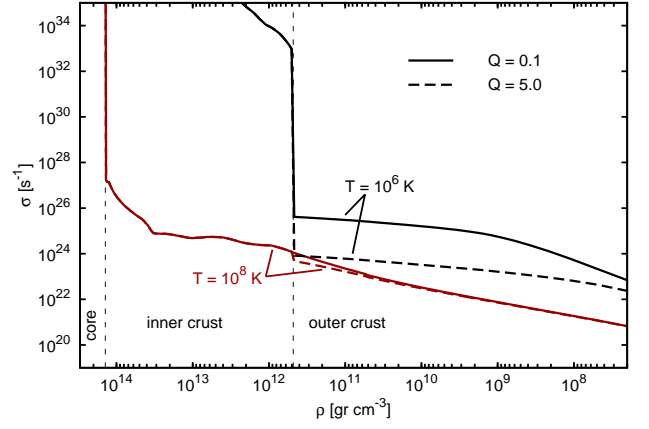}
  \caption{Electrical conductivity, $\sigma$, in the NS crust for two temperatures ($10^{6}$ and $10^{8}$~K) and two values adopted for the impurity parameter (0.1 and 5.0).}
  \label{cond}
\end{figure}

In Figure \ref{cond} we show the electrical conductivity as a function of the density in the NS interior for two different temperatures $T = 10^6$ and $10^8$~K and for two values adopted for the impurity parameter at the outer crust: $Q = 0.1$ and $5.0$. The inset at the top-right corner of the figure shows the detail of the jump of the electrical conductivity close to the ND density. Beyond the crust-core interface, the electrical conductivity is very high, several orders of magnitude higher than in the outer crust. 

In the inner crust, where we have set $Q=0$, the conductivity is strongly dependent on the temperature. For a temperature $T=10^6$~K, the conductivity presents a huge gradient, varying from $\sigma \gg 10^{35}$~s$^{-1}$ to $10^{33}$~s$^{-1}$ at the ND density. However, for a higher temperature of $10^8$~K, the gradient is less steep, varying only from $10^{27}$ to $10^{24}$~s$^{-1}$. The jump at the crust-core interface becomes smaller as the temperature decreases. 

Although at the outer crust the differences are smaller than in the inner crust, the conductivity still presents not only an important gradient, but also a strong dependence on the temperature and on the impurity parameter. For a temperature of $10^8$~K, the electrical conductivity varies from $\sim$10$^{24}$~s$^{-1}$ at the ND density to $10^{20}$~s$^{-1}$ at the NS surface. In this case, considering the higher impurity value, $Q=5.0$, a small jump in the conductivity is obtained at the ND point and, for densities $\rho < 10^{10}$~g~cm$^{-3}$, the conductivity is independent of the impurities. For a lower temperature, $T=10^6$~K, the conductivity at the outer crust behaves in a very different manner. When the impurity content is low ($Q=0.1$), $\sigma$ varies from $3 \times 10^{25}$~s$^{-1}$ at the ND density to $\sim$10$^{21}$~s$^{-1}$ at the NS surface, staying 1--3 orders of magnitude higher than for $T = 10^8$~K throughout the outer crust. For a high impurity content ($Q=5.0$), the conductivity is at least one order of magnitude lower for $\rho \gtrsim 10^8$~gr~cm$^{-3}$, becoming similar to the low-impurity case close to the NS surface. At low temperatures, the impurities become almost the only effective target for the electron scattering-on processes, and thus the jump at the ND density grows strongly as the NS cools down.

\section{Numerical results}

To solve the induction equation for the Stokes function (\ref{eq:stokes1}), we used an operator-splitting method that combines an implicit Crank-Nicolson scheme for the diffusive term and an explicit upwind method for the advective one. After several tests on the spatial step size, we fixed a grid of 200 points from the NS surface, $r=R_{\rm NS}$, to $r=10$~km, which we set $\sim$0.75~km inside the NS core, to satisfy the inner boundary condition. For the diffusive term we adopted an incremental time step $\Delta t = t / 200$. In each diffusive time step, we solved the advective term dividing this time interval into shorter ones to attain for the CFL condition \citep{cfl1928}. To calculate the diffusive term, our numerical code invokes the conductivity routines from \cite{potekhin1999} to obtain the electrical conductivity, $\sigma$, in each point of the spatial grid. Density, $\rho$, and composition, ($A$, $Z$), are taken from the equation of state and the temperature, $T$, from the thermal profiles (see Section \ref{thermal_profiles}). The impurity content, $Q=Z_{\rm imp}^2$, and the accretion rate, $\dot{M}$, are free parameters in our model.

\begin{figure}
  \centering
  \includegraphics[width=5cm,angle=-90]{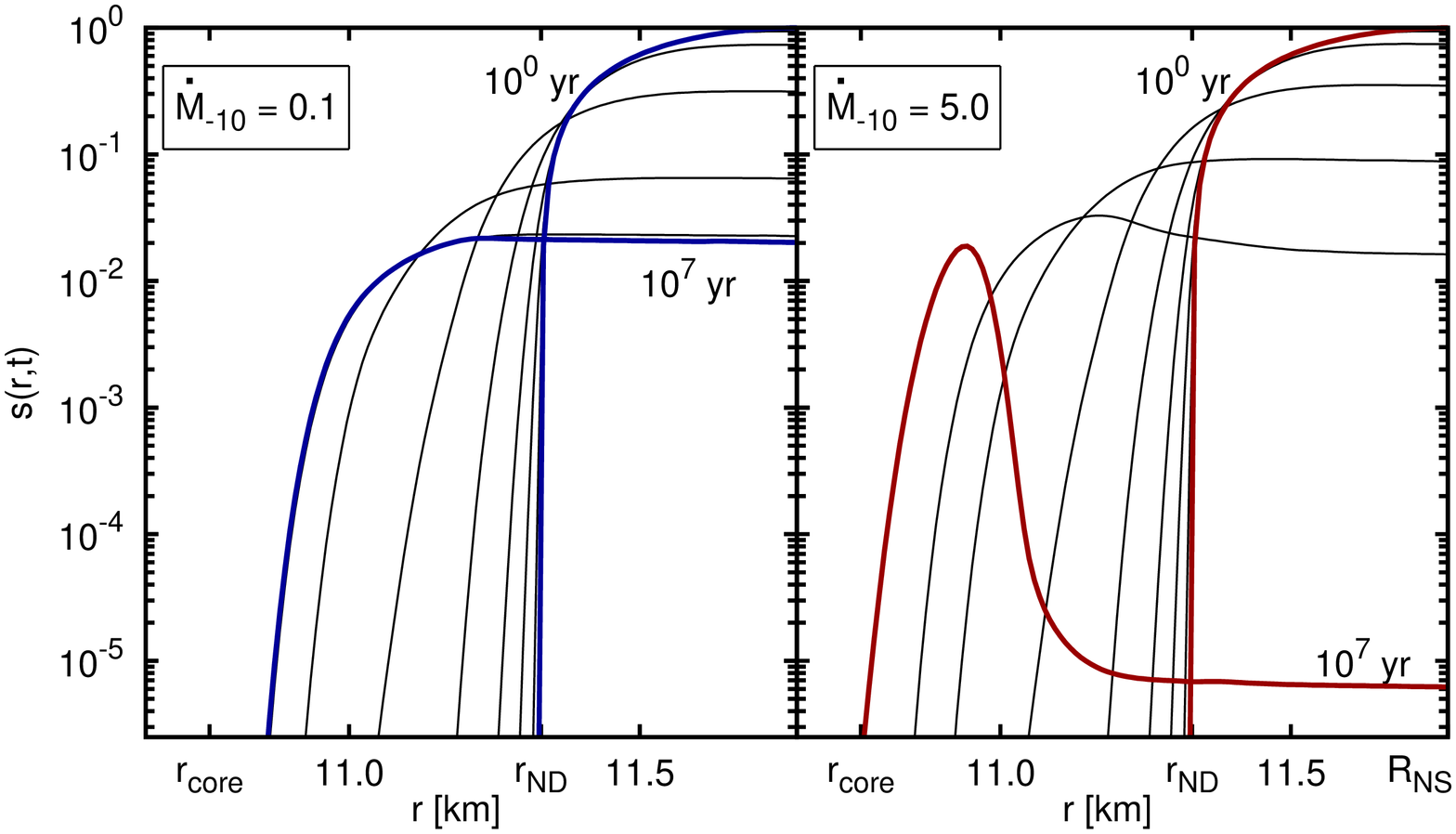}
  \caption{Evolution of the Stokes function at the crust of the LM model. Each panel corresponds to a different accretion rate. In both panels, each curve (from top to bottom at the NS surface $R_{\rm NS}$) corresponds to $t=$~1, 10$^1$, 10$^2$, 10$^3$, 1$0^4$, 10$^5$, 10$^6$ , and 10$^7$~yr. In all cases, we use $Q=5$ and {\it no SF} model.}
  \label{profileLM}
%\end{figure}
%\begin{figure}
  \centering
  \includegraphics[width=5cm,angle=-90]{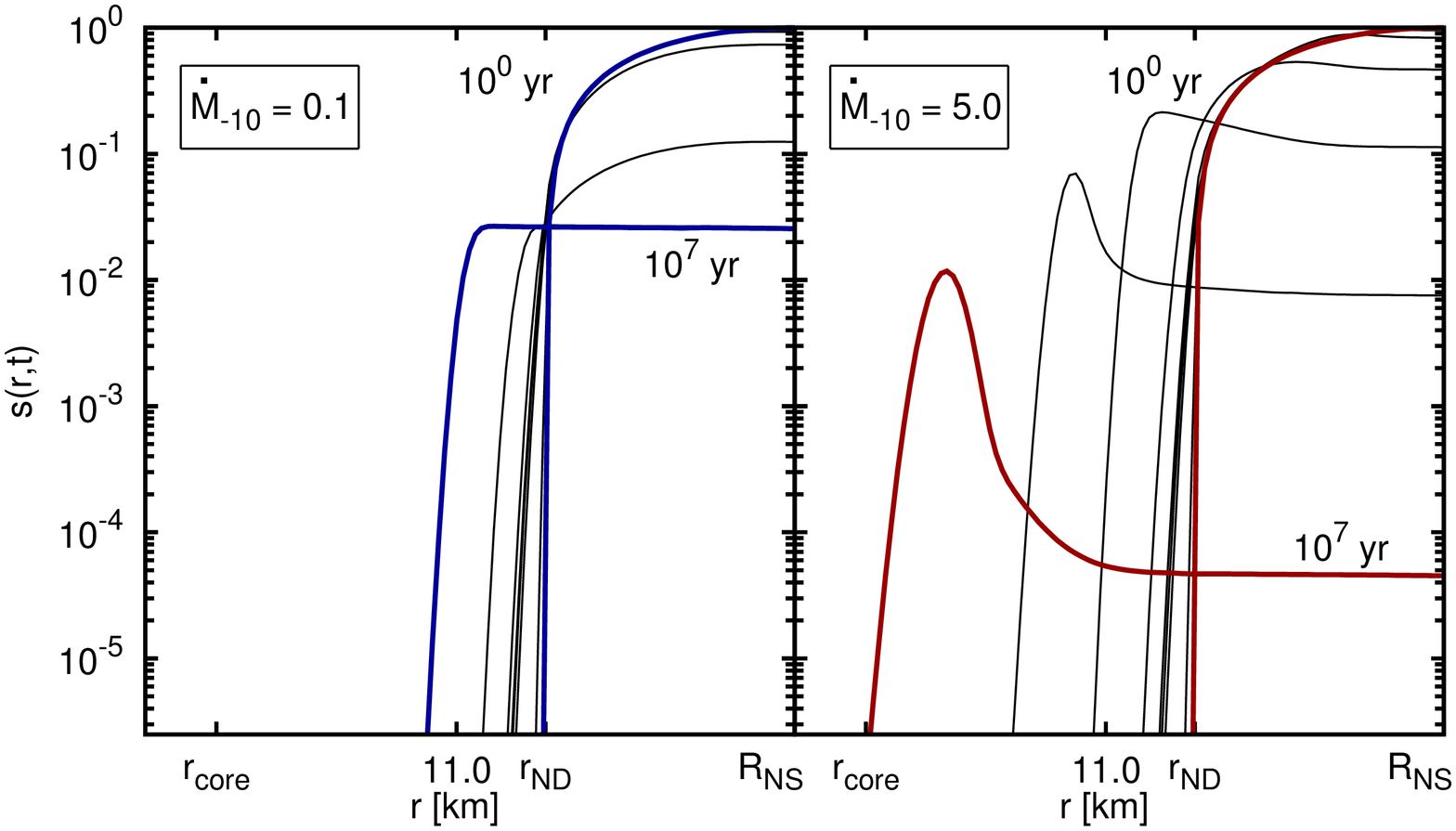}
  \caption{Idem as Figure \ref{profileLM} for HM model.}
  \label{profileHM}
\end{figure}

First we present the evolution of the Stokes' profile in the crust for the LM (Figure \ref{profileLM}) and HM (Figure \ref{profileHM}) models, considering four different values for the accretion rate, $\dot{M}_{-10}$, which is in units of $10^{-10}$~M$_{\odot}$~yr$^{-1}$. At $t=1$~yr, the Stokes profile is almost identical to the initial distribution $s(r,t=0)$ (see Eq. \ref{cond_ini}). At the first stage ($t < 10^5$~yr), the magnetic field is rearranged to the inner crust, where the conductivity is higher than in the outer crust, being reduced at most by a factor of 10 at the surface, $R_{\rm NS}$. During this first stage, the Stokes profiles in
all cases are increasing functions of the radius, as the initial distribution. For accretion rates $\dot{M} < 10^{-10}$~M$_{\odot}$~yr$^{-1}$, the decay becomes uniform over the crust, and the Stokes function reaches values $\sim$10$^{-2}$ at the NS surface on timescales $\sim$10$^7$~yr. When the effects caused by the advective term are similar to the diffusive ones, the advection of the magnetic
field through the NS interior for accretion rates $\dot{M} \gtrsim 10^{-10}$~M$_{\odot}$~yr$^{-1}$ becomes strong enough to produce a burying process, leading to significantly lower values at the surface than in the inner crust, where the conductivity is several orders of magnitude higher.

\begin{figure}
  \centering
  \includegraphics[width=8.2cm, angle=-90]{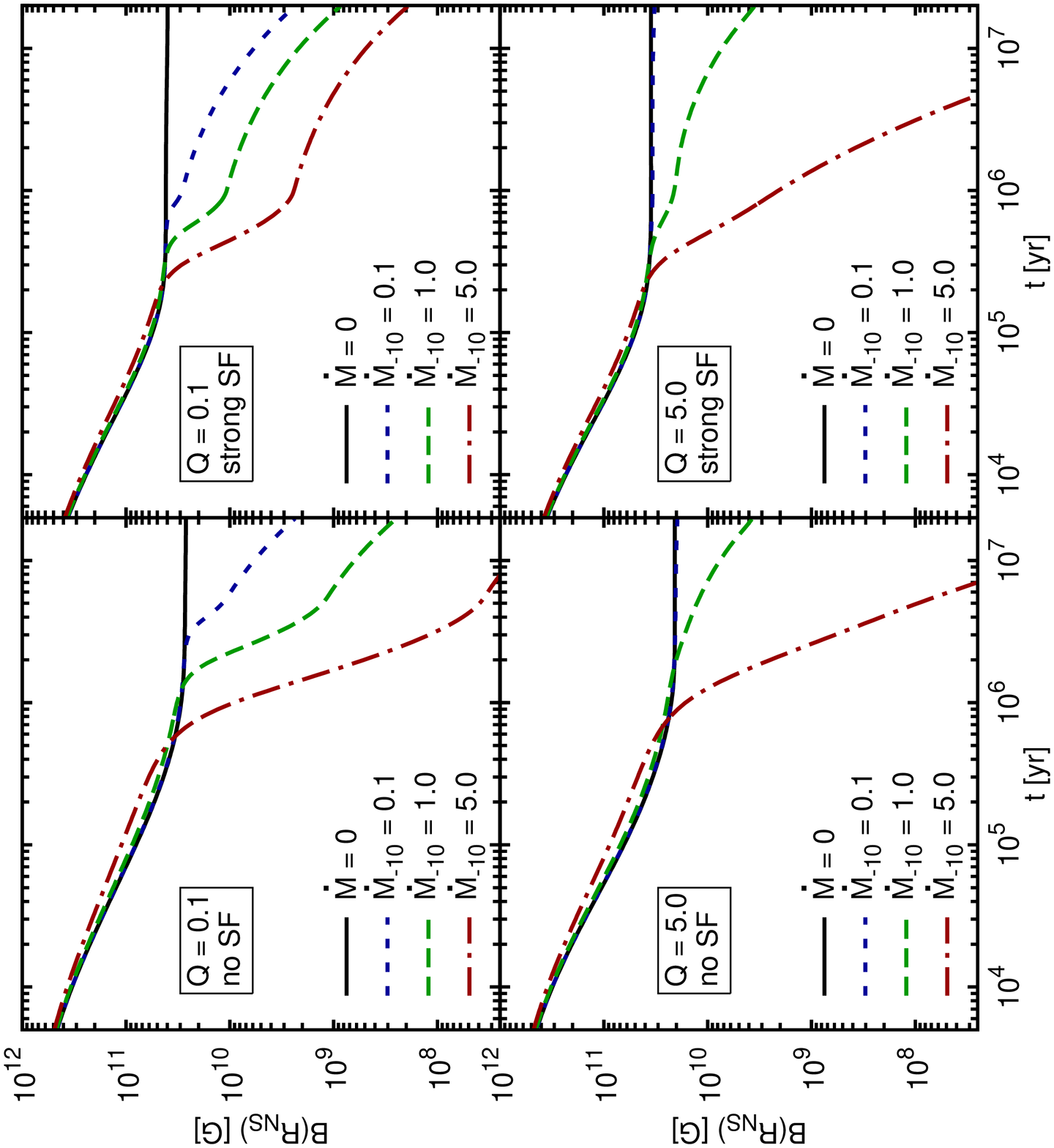}
  \caption{Surface magnetic field evolution for LM model obtained
by varying the impurity Q and cooling model.}
  \label{B_evol_LM}
%\end{figure}
%\begin{figure}
  \centering
  \includegraphics[width=8.2cm, angle=-90]{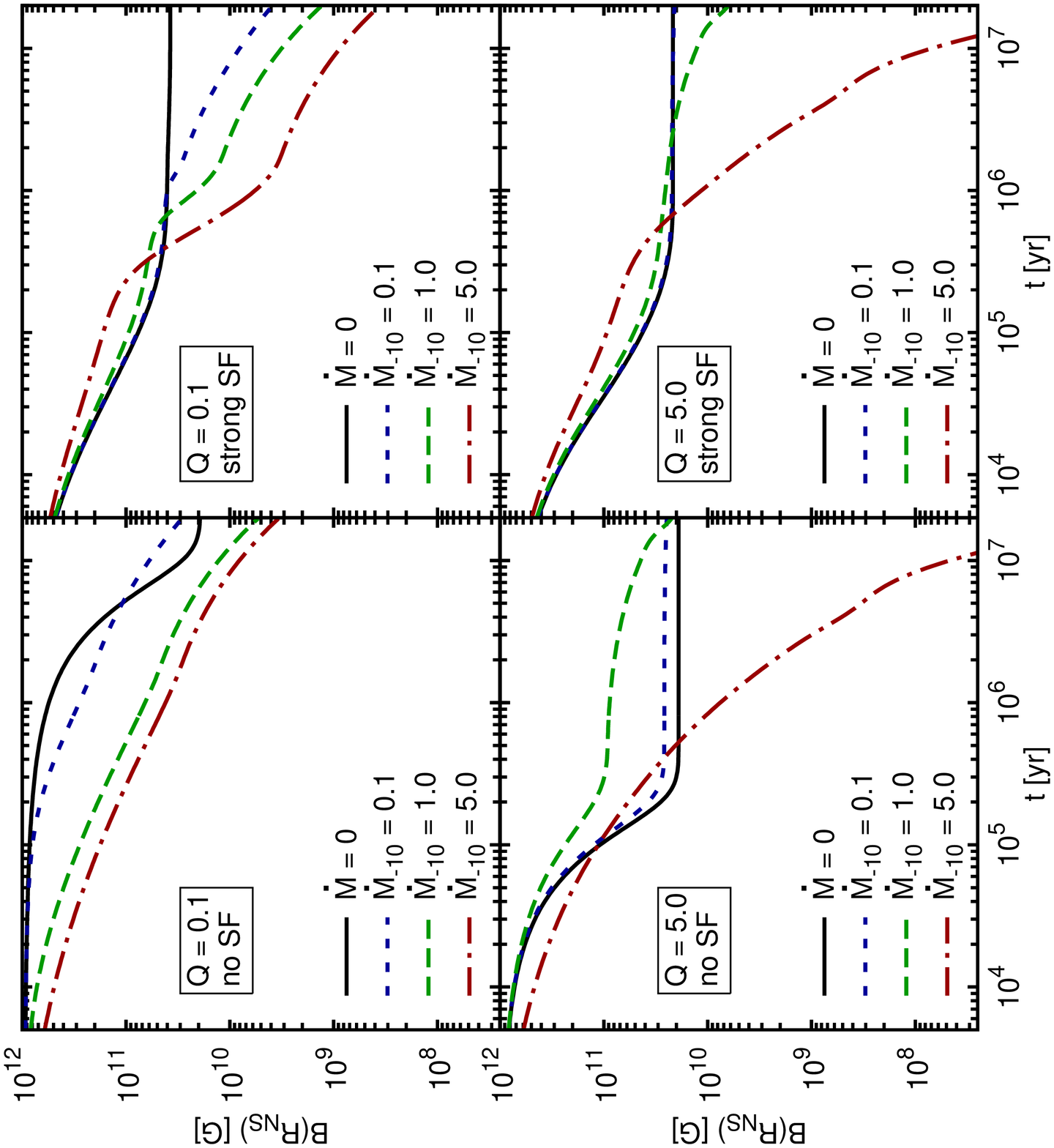}
  \caption{Idem as Figure \ref{B_evol_LM} for HM model.}
  \label{B_evol_HM}
\end{figure}

We show in Figures \ref{B_evol_LM} and \ref{B_evol_HM} the surface magnetic field evolution for the LM and HM models as a function of the accretion rate, $\dot{M}$, obtained for both superfluidity scenarios ({\it no SF} and {\it strong SF}) and for two different impurity contents at the crust ($Q=0.1$ and $Q=5.0$).

For the LM model (Figure \ref{B_evol_LM}), for the first stage ($t < 10^5$~yr), while the star is still hot ($T \gtrsim 10^8$~K), the magnetic field decay is similar in all cases, because the evolution is governed by the diffusive process. After the star cools down, the accretion process starts to dominate the decay at the surface because of the advection of the field into deeper interior layers. Thus, decay-curves separate themselves in terms of each $\dot{M}$ value. Moreover, after $t \sim 10^6$~yr, the impurity content starts to play an important role as well. For a low impurity ($Q=0.1$), when the temperature is low enough ($T \la 10^6$~K), the conductivity in the outer crust becomes $\sim$2 orders of magnitude higher than for a high-impurity content ($Q=5.0$). Thus the magnetic field is reduced faster for $Q=0.1$, because the advective process becomes more significant when the diffusion dims. For $Q=5.0$, the magnetic field is smoothly diffused in the outer crust, avoiding the burying to the interior. However, for a higher accretion rate, $\dot{M} \sim 5 \times 10^{-10}$~M$_{\odot}$~yr$^{-1}$, the magnetic field on the surface always decays below $10^8$~G in less than $2 \times 10^7$~yr. The opposite occurs if the accretion process is turned off ($\dot{M}=0$). In this case, after the magnetic field reaches the crust-core interface ($t\sim10^5-10^6$~yr), where the conductivity is extremely high, the Stokes function stops evolving over the crust, and hence the magnetic field at the surface freezes at $\sim$2$ \times 10^{10}$~G.

In summary, our results show that the decay is always fast in young stars ($t\lesssim 10^5$~yr) of low mass as long as they are warm, in agreement with previous results found by \citet{urpinmuslinov1992} and \citet{kb1997}. We obtained slightly different timescales because we incorporate updated microphysics and thermal profiles.

For the more massive HM model (Figure \ref{B_evol_HM}), the density at the core is high enough to turn on the emission of neutrinos through the direct Urca process \citep[see the review from][]{yakovlev2001}. If no neutron superfluidity is present at the core ({\it no SF} model), the star cools down very fast and magnetic field decay is very slow during the first stage ($t \la 10^5$~yr). For a superfluid interior ({\it strong SF} model), the behavior of the magnetic field evolution is found to be very similar to the LM case instead. However, in the HM case, the impurity parameter becomes crucial. Only if the impurity content is high ($Q=5.0$)
does the magnetic field decay, reaching $B \la 10^8$~G in $t \la 10^7$~yr, while if the impurity content is low ($Q=0.1$), the field remains almost a factor of 10 above this value. 

Up to this point, the results were obtained assuming that the magnetic field was expelled to the outer crust during the NS formation by fixing the $r_{\rm i}$ parameter of the initial Stokes function (see Eq. \ref{cond_ini}) to the ND point ($\rho_{\rm i} = \rho_{\rm ND}$). To study the dependence on the initial magnetic field distribution, we calculated the magnetic field evolution considering different values for this free parameter.

\begin{figure}
  \centering
  \includegraphics[width=5cm,angle=-90]{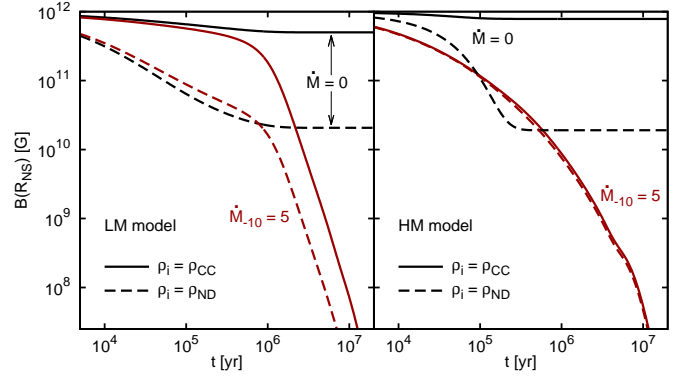}
  \caption{Surface magnetic field evolution for LM (left panel) and HM (right panel) models for two different values of the initial magnetic field expulsion: at $\rho_{\rm i} = \rho_{\rm CC} \approx 1.3 \times 10^{14}$~gr~cm$^{-3}$ (solid lines) and at $\rho_{\rm i} = \rho_{\rm ND} = 4.2 \times 10^{11}$~gr~cm$^{-3}$ (dashed lines). Two extreme accretion scenarios are compared: $\dot{M}_{-10} = 5$, for a high accretion rate, and $\dot{M}_{-10} = 0$, for no accretion (purely diffusive case). Here we set $Q=5.0$ and {\it no SF} case.}
  \label{rp_compare}
\end{figure}

In Figure \ref{rp_compare} we present the results obtained for both the LM and HM models, using two extreme cases for the initial magnetic field distribution. We compare the results shown in the left-bottom panels of Figures \ref{B_evol_LM} and \ref{B_evol_HM} for $\rho_{\rm i} = \rho_{\rm ND}$ with those found assuming a deeper initial distribution, $\rho_{\rm i} = \rho_{\rm CC} \approx 1.3 \times 10^{14}$~gr~cm$^{-3}$, which reaches the crust-core interface. In this last case, when $\dot{M} = 0$, the magnetic field remains almost constant for $10^7$~yr because of the low resistivity of the inner crust. In contrast, for high accretion rates ($\dot{M} = 5 \times 10^{-10}$~M$_{\odot}$~yr$^{-1}$) the trend of the two initial distributions is very similar. 

Under the assumptions made, we found that magnetic field decay timescales are always strongly dominated by the accretion rate. In general, we found that for $\dot{M} \ga 10^{-10}$~M$_{\odot}$~yr$^{-1}$ the magnetic field can decrease to $B\la10^8$~G in $\sim$10$^7$~yr. Furthermore, at these timescales, the field strength is found to depend on the NS mass as well through the crust thickness and the resulting thermal evolution. This thermal evolution makes the impurity content also important. In this sense, for the HM model we found that only for a high impurity content a magnetic field can be reduced by up to four orders of magnitude in less than $10^7$~yr. 

It is important to note that the impurity parameter is believed to be also strongly related to the accretion process that can be characterized through the accretion rate and total accreted mass. Nevertheless, it is not clear yet how these three parameters are related.

\section{Discussion of a jet-launching scenario for SFXTs}

In the past decade, the number of known HMXBs has grown enormously, mainly because of the surveys of the Galactic Plane carried out by the {\it INTEGRAL} satellite. In particular, several systems consisting of a compact object (a black hole or a neutron star) and a high-mass supergiant star (of O or B spectral type) have been detected. These stars have strong winds \citep[with mass losses of the order of $10^{-6}$~M$_{\odot}$~yr$^{-1}$ and velocities $v_w \sim 1000-2000$~km~s$^{-1}$,][]{vink2000}, hence X-ray emission is produced by the accretion of the wind by the compact object. Some HMXBs show transient flaring activity in X-rays, with high dynamic ranges (3--5 orders of magnitude) and short timescales (from a few hours to days) \citep{sguera2005,negueruela2006}. The distance obtained from their optical/IR counterparts implies typical X-ray luminosities $L_{\rm X}\sim10^{36}$~erg~s$^{-1}$ in outburst and $L_{\rm X}\sim 10^{32}$~erg~s$^{-1}$ in quiescence. Most of the time, however, the sources are in an intermediate state of $L_{\rm X}\sim10^{33}-10^{34}$~erg~s$^{-1}$, the product of residual accretion onto the compact object \citep{sidoli2008}.

In some of these systems outburts are regular and coincident with the orbital period of the binary system, but in most cases no period has been found and a random behaviour prevails. To explain this phenomenology, \citet{intzand2005} proposed that the transient X-ray flaring is produced by the accretion of clumps originating from a highly structured wind from the supergiant companion \citep[see e.g.,][]{runacres2005,owocki2006}. This idea was explored by \citet{walterzurita2007} and later by \citet{negueruela2008}, leading to a thorough clumpy wind model developed by \citet{ducci2009}. To explain the periodic outbursts from IGR J11215--5952, \citet{sidoli2007a} proposed that the regular flaring activity might be generated when the compact object crosses a slow and dense disk-like wind component, twice per orbital period, which proposition was based on an accretion scenario. Another model proposed by \citet{grebenevsunyaev2007} suggested that high-energy transient emission might be produced in these systems by the interaction of the stellar wind with the magnetosphere of the NS, in a so-called gated mechanism. \citet{bozzo2008} showed that this is possible for a magnetic field of the NS of the order of $B\sim10^{14}-10^{15}$~G, as in the so-called magnetars. A cyclotron emission line at $3.3$~keV in the SFXT IGR J18483--0311 allowed infering a magnetic field $B\sim10^{11}$~G, assuming an electron origin. Nevertheless, a magnetar cannot be ruled out if the line is caused by protons, for which $B\sim5\times10^{14}$~G \citep{sguera2010}.

Three SFXTs (and one candidate) are spatially correlated with unidentified transient $\gamma$-ray sources detected by {\it EGRET} and/or {\it AGILE} satellites, as mentioned in the introduction. They are IGR J20188+3647/AGL J2022+3622, AX J1841.0--0536/3EG J1837--0423, IGR J11215--5952/3EG J1122-5946 \citep{sguera2009}, and  IGR J17354--3255/AGL J1734--3310 \citep{sguera2011}. No {\it Fermi} source has been found to our knowledge, but this is not surprising because of the transient character of the phenomena and the survey-mode operation of the instrument\footnote{The source-detection step of the latest {\it Fermi} catalogue \citep{nolan2012} was only applied to the data from the full 24-month time interval of the data set. No search for transient sources that may have been bright for only a small fraction of the 2-year interval was systematically implemented so far. The {\it a priori} probability of detection of moderate transient sources with very short duty-cycles, such as those that might be associated with SFXTs, is very low.}. If the $\gamma$-ray emission is real, the spatial association opens the possibility of a common origin of the emission at different wavelengths. 

We briefly discuss these sources individually below. 

\subsubsection*{AX J1841.0--0536}

AX J1841.0--0536 is a transient X-ray pulsar ($P_{\rm s}=4.7$~s) detected by the {\it ASCA} satellite in 1994 and 1999 \citep{bamba2001}, while displaying X-ray flares of a factor of $\sim$10 on timescales of $\sim$1 hour. A {\it Chandra} observation allowed determining
its coordinates with precision \citep{halperngotthelf2004}. This enabled \cite{halpern2004} to find an optical/IR counterpart of the system, which is a supergiant star of spectral type B1 Ib \citep{nespoli2008}. The source was also detected during an outburst by {\it INTEGRAL} and {\it Swift}, and finally confirmed as a member of the SFXT class after the deep study of \cite{romano2011}.

In the third {\it EGRET} catalogue, 3EG J1837--0423 is a transitory point source, and although the spatial correlation with AX J1841.0--0536 is ambiguous, the absence of another hard X-ray source in the region suggests a physical relation between them \citep{sgueraromero2009}.

\subsubsection*{IGR J11215--5952}

This source was discovered with {\it INTEGRAL} by \citet{lubinski2005}. It is associated with a B0.7 supergiant counterpart at a distance of 8 kpc \citep{negueruela2007}. Pulsed emission of $P_{\rm s} = 187$~s \citep{swank2007} showed that the compact object in this SFXT is an NS. Moreover, this is the first periodic member of the class, showing regular outbursts every $\sim$165~days according to \cite{romano2009}, with typical X-ray luminosities of $L_{\rm X} \sim 5 \times 10^{36}$~erg~s$^{-1}$, and long quiescent states of $L_{\rm X} \sim 10^{33}$~erg~s$^{-1}$.

The unidentified {\it EGRET} source EGR J1122--5946 is well spatially correlated with SFXT IGR J11215--5952 \citep{sguera2009}. In addition, the absence of other counterparts in the soft $\gamma$ band suggests the possibility of a physical association between them. However, {\it EGRET} observations do not allow to confirm whether the source is constant or transient.

\subsubsection*{IGR J20188+3647}

In 2004, {\it INTEGRAL} discovered the transient X-ray source IGR J20188+3647. Its X-ray flaring properties, with timescales of $\sim$1 hour, resemble those of SFXTs, and hence it is considered as an SFXT candidate \citep{sguera2006b}. 

An unidentified $\gamma$-ray source, named AGL J2022+3622, was observed by {\it AGILE} \citep{chen2007} in the same region. It is a variable source in the MeV band, seen active for approximately only one day. IGR J20188+3647 is the only hard X-ray source inside the positional error circle of this {\it AGILE} source \citep{sguera2009}.

\subsubsection*{IGR J17354--3255}

IGR J17354--3255 is a hard X-ray transient discovered by {\it INTEGRAL} \citep{kuulkers2006,kuulkers2007}. With an orbital period of 8.4~days \citep{dai2011}, the source presents flares with X-ray luminosities $L_{\rm X} \gtrsim 10^{36}$~erg~s$^{-1}$ with a dynamic range $\gtrsim$20, typical of intermediate SFXTs \citep{sguera2011,bozzo2012}. 
Recently, \citet{coleiro2013} found the near-IR counterpart of the system, an 09Iab supergiant, thus confirming its SFXT classification.

This {\it INTEGRAL} source is the only hard X-ray source unambiguously located inside the error circle of the unidentified source AGL J1734--3310 \citep{sguera2011,sguera2013}, which is a transient MeV/GeV source detected in outburst by the {\it AGILE} satellite \citep{bulgarelli2009}. Here, the possible association is based not only on the spatial correlation, but also on the temporal behaviour of the two sources at different wavelengths. However, based on observations performed in the soft X-ray band by the {\it Swift}/XRT, \cite{ducci2013} suggest that IGR J17354-3255 is an almost persistent HMXB. The authors argued against the SFXT nature of this source, proposing an eclipse origin for its variability.

It is important to remark that the only observational evidence available at present to support a possible physical association of these sources is a certain spatial, and sometimes temporal, correlation. It is clear that long-time observations both in $\gamma$ and X-ray bands are necessary to establish a physical connection among these sources.  

As was argued by \citet{sgueraromero2009}, a promising approach to explaining the production of relativistic particles capable of generating a transient $\gamma$-ray source, under the particular conditions imposed by the SFXTs, is the formation of a transient jet powered by a magnetic tower. Such an outflow could carry away a considerable fraction of the accreting material \citep{kato2007}. In the magneto-centrifugal model of jet formation, the ejection of matter is only possible if the material of the inner part of the accretion disk can reach distances of $\sim$40 gravitational radii \citep{kato2004}. Thus, the Alfv\'en radius, $R_{\rm A}$, cannot be much larger than the NS radius, $R_{\rm NS}$, implying that the magnetic field of the NS must be weak enough to allow the matter penetration. In this sense, a simple basic condition for jet formation in accreting NSs was obtained by \citet{massikauf2008}, who showed that the magnetic field at the surface of the NS must be $B\lesssim10^8$~G.

Although jet emission in accreting NSs was already observed in some LMXBs \citep[e.g. Cir X-1,][]{fender1998}, this phenomenon was not detected in HMXBs yet (but see the discussion about \mbox{LS I +61 303}, \citet{romero2007}, for which jet models have been proposed, \citet{boschramon2006}). Our numerical results suggest that if a significant fraction of the strong wind emitted by the donor star can be accreted by the NS, a magnetic field evolution from a typical $B=10^{12}$~G surface field to $B\la 10^8$~G might be possible on timescales $t \sim 10^7$~yr, allowing for jet formation.

\begin{figure}
  \centering
  \includegraphics[height=8.8cm,angle=-90, trim=0 0 0 0,clip]{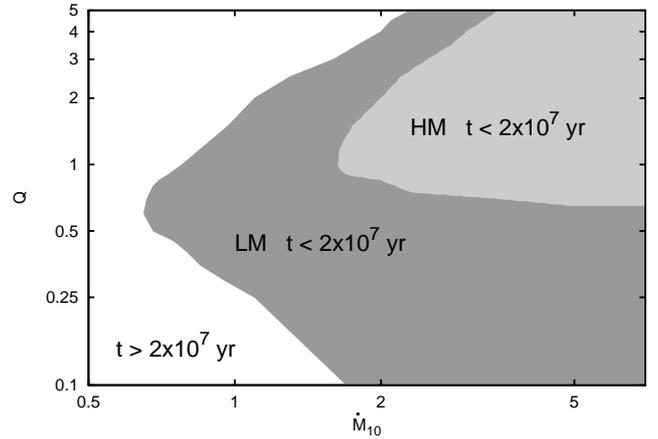}
  \caption{Timescale, $t$, for magnetic field decay from $10^{12}$~G to $\sim$10$^{8}$~G, as a function of the impurity content, $Q$, and the accretion rate, $\dot{M}_{-10}$, in units of $10^{-10}$~M$_{\odot}$~yr$^{-1}$, for both LM and HM models, in the {\it no SF} case.}
  \label{plot1}
\end{figure}

In Figure \ref{plot1} we summarize the timescale for magnetic field decay from a typical initial pulsar-strength, $B=10^{12}$~G to $B\la$10$^8$~G, which is necessary for jet formation, as a function of the impurity content in the crust, $Q$, and the accretion rate, $\dot{M}$, for LM and HM models (considering no neutron superfluidity in the core, {\it no SF}). In the plot, it can be observed that if $\dot{M} < 6 \times 10^{-11}$~M$_{\odot}$~yr$^{-1}$, the magnetic field cannot decrease four orders of magnitude in $t < 2 \times 10^7$~yr. In contrast, if $\dot{M} > 1.7 \times 10^{-10}$~M$_{\odot}$~yr$^{-1}$ , the magnetic field may decay by more than four orders of magnitude at the surface for $Q>0.65$ in the HM model, while in the LM model, such a decay might be possible for all the explored values of the impurity content. In both models, for $0.6 \la Q \la 1.2$, a decrease of four orders of magnitude might be possible for a much wider range of $\dot{M}$.

In our approach to the problem we assumed important constraining hypotheses that we discuss now.  First, the magnetic field has a fixed geometry: a dipolar configuration. We have neglected higher-order terms that cannot evolve. However, the local reorganization of the field has been found to have important effects on a short timescale and for strongly magnetized sources like magnetars for which $B\sim 10^{14}$~G \citep{pons2009, pons2012}. Although including higher-order terms in the field description is more realistic, the long-term evolution would be dominated by the global geometry of the field and the ohmic diffusion and convection caused by the accreted material.

Second, the accretion process was assumed to be spherical and with constant mass rate, neglecting any redirection of the material through the magnetic field poles and any stellar wind dynamics over $\sim$10$^7$~yr. For fields $B \sim 10^8 - 10^{12}$~G and typical wind accretion rates, the magnetic energy density outside the star is much higher than the energy density of the accreting matter, and thus matter will flow onto the star's surface in narrow channels. A more realistic scenario should include a two-dimensional treatment for which the material is accumulated at the magnetic polar caps with a local accretion rate as the relevant parameter that varies according to the evolution of the companion star. Although the modification of the results shown here under a two-dimensional columnar accretion model are difficult to predict, qualitatively speaking, the burying of the magnetic field that dominates the evolution at the surface should also take place, and thus the net effect found in the one-dimensional spherical model might play an important role in a columnar accretion scenario as well \citep[see, for instance,][]{payne2004,payne2007,lovelace2005}. The results shown in this work will certainly motivate future studies in this direction, and for these we mention that such theoretical efforts are most likely not in correspondence with other uncertainties of the problem, such as the observationally inferred accretion rates and the stellar dynamics themselves.

Another approximation we adopted was that the crust is isothermal and its evolution does not consider possible heat deposition due to the accretion process. Nevertheless, in the long-term evolution the results might be more affected by the crust composition; therefore we have considered a modified accreted equation of state for the thermal profiles, as in \citet{aguilera2008}. 

Finally, the impurity content has been set as a constant parameter through the outer crust, and with no relation with the intensity of the accretion process (total accreted mass and/or accretion period) because no quantitative relation between them has been found so far; recent studies of quiescent thermal emission of LMXBs indicate that values of $Q\sim 1$ are required to adjust data for two sources whose accretion rates are in the range $1-5 \times 10^{-10}$~M$_{\odot}$~yr$^{-1}$ \citep{turlioneaguilera2013}. 

\section{Summary}

We studied necessary conditions for jet formation in binary systems formed by a neutron star that undergoes an induced magnetic field decay due to the accretion of the wind produced by a high-mass stellar companion. We used a numerical model that incorporates state-of-the-art microphysics involving the electrical conductivity at the neutron star crust \citep{potekhin1999,chugunov2012} and consistent thermal evolution profiles \citep{aguilera2008}. Our treatment took into account diffusion of the currents in the neutron star crust and advection of the field as a direct consequence of the matter accretion onto the surface. A strong limitation of our model is that it considers spherical accretion instead of columnar accretion. We focused, however, on qualitative aspects of the evolution of the surface magnetic field - such as burying
- that are expected to be important in the accretion area in a two-dimensional treatment as well. They may  also play a fundamental role in jet formation.

The results indicate that the timescale of the magnetic field decay strongly depends on the accretion rate, which is less sensitive to the impurity parameter $Q$ and indirectly, through the thermal profile, to the neutron star mass and to the superfluidity state of the core. If a substantial fraction of the wind emitted by the companion is accreted by the neutron star, for instance, $\dot{M}\gtrsim 10^{-10}$~M$_{\odot}$~yr$^{-1}$, the magnetic field can decay from a typical initial value $B=10^{12}$~G to $B \la 10^8$~G on a timescale $t \sim 10^7$~yr. Hence, it could be possible for neutron stars to launch jets in HMXBs. 

Our results are important for models of SFXTs and gamma-ray binaries such as \mbox{LS I +61 303} or \mbox{LS 5039}, where non-thermal phenomena are observed, and accretion rates $\sim$10$^{-10}$~M$_{\odot}$~yr$^{-1}$ have been deduced from simulations \citep[see][]{romero2007,owocki2011}. In particular, hard fast X-ray transients might form a new class of Galactic MeV emitters, as was argued by \cite{sgueraromero2009} for the SFXT AX J1841.0--0536. 

Deep observations in $\gamma$-rays with {\sl Fermi} and {\sl AGILE} satellites of the four candidate sources presented here will be important to confirm the existence of transient relativistic particle injection when, for instance, large clumps are accreted from the stellar wind, allowing for the ejection of jets. Higher energy dedicated observations with MAGIC II and HESS II might also help to constrain the level of non-thermal radiation during X-ray flares and to determine the cut-off and maximum energy of the relativistic particles.  Radio interferometric observations with high angular resolution will be necessary to detect the relativistic outflows. The detection of a transient jet in an HMXB with NSs would provide indirect evidence for our predictions of the magnetic field decay.

\begin{acknowledgements}
We thank Ulrich Geppert and the referee for constructive comments. The research of DNA was partially supported by CONICET and PIP-2011-00170 (Argentina). GER is supported by CONICET (PIP-2010-0078) and ANPCyT (PICT 2012-00878), as well as by Spanish Ministerio de Ciencia e Innovaci\'on (MICINN) under grant AYA2010-21782-C03-01.
 
\end{acknowledgements}


\begin{thebibliography}{}

\bibitem[Aguilera et al.(2008)]{aguilera2008} Aguilera, D.~N., Pons, J.~A., \& Miralles, J.~A.\ 2008, \aap, 486, 255 

\bibitem[Bamba et al.(2001)]{bamba2001} Bamba, A., Yokogawa, J., Ueno, M., Koyama, K., \& Yamauchi, S.\ 2001, \pasj, 53, 1179 

\bibitem[Bisnovatyi-Kogan \& Komberg(1974)]{bisno1974} Bisnovatyi-Kogan, G.~S., \& Komberg, B.~V.\ 1974, \sovast, 18, 217 

\bibitem[Bosch-Ramon et al.(2006)]{boschramon2006} Bosch-Ramon, V., Paredes, J. M., Romero, G. E., \& Rib, M. 2006, \aap, 459, L25 

\bibitem[Bozzo et al.(2008)]{bozzo2008} Bozzo, E., Falanga, M., \& Stella, L.\ 2008, \apj, 683, 1031 

\bibitem[Bozzo et al.(2012)]{bozzo2012} Bozzo, E., Pavan, L., Ferrigno, C., et al.\ 2012, \aap, 544, A118 

\bibitem[Bulgarelli et al.(2009)]{bulgarelli2009} Bulgarelli, A., Gianotti, F., Trifoglio, M., et al.\ 2009, The Astronomer's Telegram, 2017, 1 

\bibitem[Chen et al.(2007)]{chen2007} Chen, A., Vercellone, S., Giuliani, A., et al.\ 2007, The Astronomer's Telegram, 1308, 1 

\bibitem[Chugunov(2012)]{chugunov2012} Chugunov, A.~I.\ 2012, Astronomy Letters, 38, 25 

\bibitem[Coleiro et al.(2013)]{coleiro2013} Coleiro, A., Chaty, S., Zurita Heras, J.~A., Rahoui, F., \& Tomsick, J.~A.\ 2013, arXiv:1310.0451 

\bibitem[Courant et al.(1928)]{cfl1928} Courant, R., Friedrichs, K., \& Lewy, H.\ 1928, Mathematische Annalen, 100, 32

\bibitem[D'A{\`i} et al.(2011)]{dai2011} D'A{\`i}, A., La Parola, V., Cusumano, G., et al.\ 2011, \aap, 529, A30 

\bibitem[Douchin \& Haensel(2001)]{douchin2001} Douchin, F., \& Haensel, P.\ 2001, \aap, 380, 151 

\bibitem[Ducci et al.(2009)]{ducci2009} Ducci, L., Sidoli, L., Mereghetti, S., Paizis, A., \& Romano, P.\ 2009, \mnras, 398, 2152 

\bibitem[Ducci et al.(2013)]{ducci2013} Ducci, L., Romano, P., Esposito, P., et al.\ 2013, \aap, 556, A72 

\bibitem[Fender et al.(1998)]{fender1998} Fender, R., Spencer, R., Tzioumis, T., et al.\ 1998, \apjl, 506, L121 

\bibitem[Geppert \& Urpin(1994)]{gu1994} Geppert, U., \& Urpin,  V.\ 1994, \mnras, 271, 490  

\bibitem[Grebenev \& Sunyaev(2007)]{grebenevsunyaev2007} Grebenev, S.~A., \& Sunyaev, R.~A.\ 2007, Astronomy Letters, 33, 149 

\bibitem[Haensel \& Zdunik(2008)]{haensel2008} Haensel, P., \& Zdunik, J.~L.\ 2008, \aap, 480, 459 

\bibitem[Halpern \& Gotthelf(2004)]{halperngotthelf2004} Halpern, J.~P., \& Gotthelf, E.~V.\ 2004, The Astronomer's Telegram, 341, 1 

\bibitem[Halpern et al.(2004)]{halpern2004} Halpern, J.~P., Gotthelf, E.~V., Helfand, D.~J., Gezari, S., \& Wegner, G.~A.\ 2004, The Astronomer's Telegram, 289, 1 

\bibitem[in't Zand(2005)]{intzand2005} in't Zand, J.~J.~M.\ 2005, \aap, 441, L1 

\bibitem[Kato et al.(2004)]{kato2004} Kato, Y., Hayashi, M.~R., \& Matsumoto, R.\ 2004, \apj, 600, 338 

\bibitem[Kato(2007)]{kato2007} Kato, Y.\ 2007, \apss, 307, 11 

\bibitem[Konar \& Bhattacharya(1997)]{kb1997} Konar, S., \& Bhattacharya, D.\ 1997, \mnras, 284, 311 
  
\bibitem[Kuulkers et al.(2006)]{kuulkers2006} Kuulkers, E., Shaw, S., Paizis, A., et al.\ 2006, The Astronomer's Telegram, 874, 1 

\bibitem[Kuulkers et al.(2007)]{kuulkers2007} Kuulkers, E., Shaw, S.~E., Paizis, A., et al.\ 2007, \aap, 466, 595 

\bibitem[Lovelace et al.(2005)]{lovelace2005} Lovelace, R.~V.~E., Romanova, M.~M., \& Bisnovatyi-Kogan, G.~S.\ 2005, \apj, 625, 957 

\bibitem[Lubi{\'n}ski et al.(2005)]{lubinski2005} Lubi{\'n}ski, P., Bel, M.~G., von Kienlin, A., et al.\ 2005, The Astronomer's Telegram, 469, 1 

\bibitem[Massi \& Kaufman Bernad{\'o}(2008)]{massikauf2008} Massi, M., \& Kaufman Bernad{\'o}, M.\ 2008, \aap, 477, 1 

\bibitem[Negueruela et al.(2006)]{negueruela2006} Negueruela, I., Smith, D.~M., Reig, P., Chaty, S., \& Torrej{\'o}n, J.~M.\ 2006, The X-ray Universe 2005, 604, 165 

\bibitem[Negueruela et al.(2007)]{negueruela2007} Negueruela, I., Smith, D.~M., Torrej{\'o}n, J.~M., \& Reig, P.\ 2007, ESA Special Publication, 622, 255 

\bibitem[Negueruela et al.(2008)]{negueruela2008} Negueruela, I., Torrej{\'o}n, J.~M., Reig, P., Rib{\'o}, M., \& Smith, D.~M.\ 2008, A Population Explosion: The Nature \& Evolution of X-ray Binaries in Diverse Environments, 1010, 252 

\bibitem[Negueruela(2010)]{negueruela2010} Negueruela, I.\ 2010, High Energy Phenomena in Massive Stars, 422, 57 

\bibitem[Nespoli et al.(2008)]{nespoli2008} Nespoli, E., Fabregat, J., \& Mennickent, R.~E.\ 2008, \aap, 486, 911 

\bibitem[Nolan et al.(2012)]{nolan2012} Nolan, P.~L., Abdo, A.~A., Ackermann, M., et al.\ 2012, \apjs, 199, 31 

\bibitem[Owocki \& Cohen(2006)]{owocki2006} Owocki, S.~P., \& Cohen, D.~H.\ 2006, \apj, 648, 565 

\bibitem[Owocki et al.(2011)]{owocki2011} Owocki, S.~P., Okazaki, A.~T., \& Romero, G.\ 2011, IAU Symposium, 272, 587 

\bibitem[Payne \& Melatos(2004)]{payne2004} Payne, D.~J.~B., \& Melatos, A.\ 2004, \mnras, 351, 569 

\bibitem[Payne \& Melatos(2007)]{payne2007} Payne, D.~J.~B., \& Melatos, A.\ 2007, \mnras, 376, 609 

\bibitem[Pons et al.(2009)]{pons2009} Pons, J.~A., Miralles, J.~A., \& Geppert, U.\ 2009, \aap, 496, 207

\bibitem[Potekhin(1999)]{potekhin1999} Potekhin, A.~Y.\ 1999, \aap, 351, 787 

\bibitem[Romano et al.(2009)]{romano2009} Romano, P., Sidoli, L., Cusumano, G., et al.\ 2009, \apj, 696, 206

\bibitem[Romano et al.(2011)]{romano2011} Romano, P., Vercellone, S., Krimm, H.~A., et al.\ 2011, arXiv:1111.0698 

\bibitem[Romero et al.(2007)]{romero2007} Romero, G.E., Okazaki, A. T,  Orellana, M., \&  Owocki, S. P. 2007, \aap, 474, 15 

\bibitem[Runacres \& Owocki(2005)]{runacres2005} Runacres, M.~C., \& Owocki, S.~P.\ 2005, \aap, 429, 323 

\bibitem[Sguera et al.(2005)]{sguera2005} Sguera, V., et al.\ 2005, \aap, 444, 221 

\bibitem[Sguera et al.(2006)]{sguera2006b} Sguera, V., Bird, A.~J., Dean, A.~J., et al.\ 2006, The Astronomer's Telegram, 873, 1 

\bibitem[Sguera(2009)]{sguera2009} Sguera, V.\ 2009, arXiv:0902.0245, Proceedings of the 7th INTEGRAL Workshop

\bibitem[Sguera et al.(2009)]{sgueraromero2009} Sguera, V., Romero, G.~E., Bazzano, A., Masetti, N., Bird, A.~J., \& Bassani, L.\ 2009, \apj, 697, 1194 

\bibitem[Sguera et al.(2010)]{sguera2010} Sguera, V., Ducci, L., Sidoli, L., Bazzano, A., \& Bassani, L.\ 2010, \mnras, 402, L49 

\bibitem[Sguera et al.(2011)]{sguera2011} Sguera, V., Drave, S.~P., Bird, A.~J., et al.\ 2011, \mnras, 417, 573 

\bibitem[Sguera(2013)]{sguera2013} Sguera, V.\ 2013, Nuclear Physics B Proceedings Supplements, 239, 76 

\bibitem[Sidoli et al.(2007)]{sidoli2007a} Sidoli, L., Romano, P., Mereghetti, S., et al.\ 2007, \aap, 476, 1307 

\bibitem[Sidoli et al.(2008)]{sidoli2008} Sidoli, L., Romano, P., Mangano, V., et al.\ 2008, \apj, 687, 1230 

\bibitem[Sidoli(2009)]{sidoli2009} Sidoli, L.\ 2009, Advances in Space Research, 43, 1464 

\bibitem[Sidoli(2011)]{sidoli2011} Sidoli, L.\ 2011, Advances in Space Research, 48, 88 

\bibitem[Swank et al.(2007)]{swank2007} Swank, J.~H., Smith, D.~M., \& Markwardt, C.~B.\ 2007, The Astronomer's Telegram, 999, 1

\bibitem[Turlione et al.(2013)]{turlioneaguilera2013} Turlione, A., Aguilera, D.~N., \& Pons, J.~A.\ 2013, arXiv:1309.3909 

\bibitem[Urpin \& Muslimov(1992)]{urpinmuslinov1992} Urpin, V.~A., \& Muslimov, A.~G.\ 1992 \mnras, 256, 261 

\bibitem[Vigan{\`o} et al.(2012)]{pons2012} Vigan{\`o}, D., Pons, J.~A., \& Miralles, J.~A.\ 2012, Computer Physics Communications, 183, 2042 

\bibitem[Vink et al.(2000)]{vink2000} Vink, J.~S., de Koter, A., \& Lamers, H.~J.~G.~L.~M.\ 2000, \aap, 362, 295 

\bibitem[Walter \& Zurita Heras(2007)]{walterzurita2007} Walter, R., \& Zurita Heras, J.\ 2007, \aap, 476, 335 

\bibitem[Yakovlev et al.(2001)]{yakovlev2001} Yakovlev, D.~G., Kaminker, A.~D., Gnedin, O.~Y., \& Haensel, P.\ 2001, \physrep, 354, 1 

\end{thebibliography}
\end{document}